# Dynamic Analysis of Regression Problems in Industrial Systems: Challenges and Solutions

Fabrizio Pastore and Leonardo Mariani

University of Milano - Bicocca
Department of Informatics, Systems and Communication
{pastore,mariani}@disco.unimib.it

**Abstract.** This paper presents the result of our experience with the application of runtime verification, testing and static analysis techniques to several industrial projects. We discuss the eight most relevant challenges that we experienced, and the strategies that we elaborated to face them.

## 1 Dynamic Analysis of Regression Problems

Industrial software systems are complex systems that must evolve quickly to remain competitive. Consider for instance the case of unmanned airborne vehicles, which might be upgraded to extend the set of missions that can be performed, to support new devices that might be installed on the vehicles, to fix bugs, and to eliminate inefficiencies. Although upgrades are designed to improve software systems, they may also expose systems to *regression problems*, that is the changes may introduce unwanted side-effects that break existing functionalities.

Regression problems are known to be tricky to detect and expensive to fix because they are the result of unexpected interactions between the changes and the existing functionalities. In particular, *localizing* and *understanding* the causes of regression failures can be extremely challenging [16, 18].

In the context of the European project PINCETTE [3], we addressed the automatic detection of regression problems by integrating *runtime verification*, *testing*, and *static analysis*, as shown in Fig. 1. The analysis starts from a *base* software version and an *upgraded* software version; the latter extending the base version with new features while introducing some regression faults. Our approach first *automatically discovers relevant behavioral properties for the base version* of the software by collecting *traces* (Step 1, Fig. 1) and mining *properties* from traces (Step 2) [14, 15]. Since mined properties might be imprecise, it uses *static analysis* (e.g., model checking) to eliminate inaccurate properties and keep only the *true properties* that have been proved to be correct for the base version (Step 3) [15]. For example, this approach may automatically discover that the speed of a vehicle cannot be negative and discard a property that allows for negative speed values.

Our approach continues by *automatically determining the obsolete properties, that is the properties that have been intentionally invalidated by the change.*

This is done by executing the passing test cases designed for the upgraded system (Step 4) while verifying the true properties at runtime (Step 5) [15]. Since property violations are detected by running passing test cases, the violations indicate behaviors that have been intentionally changed by the upgrade. For example, a vehicle may allow for negative speeds once the capability to move backward has been introduced, intentionally violating properties that state that the speed must be positive or 0. The remaining properties are the *up-to-date properties*, that is the properties that hold on the base version and have not been intentionally invalidated by the upgrade.

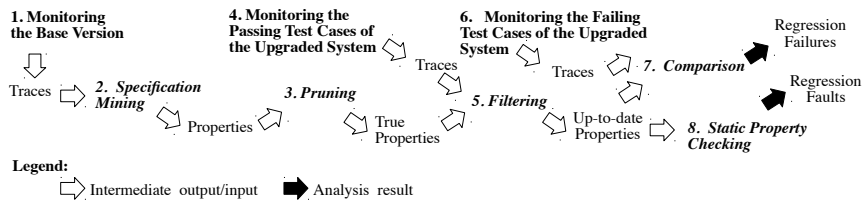

**Fig. 1.** Detection and Analysis of Regression Faults.

Finally, if there is any failing test case designed for the upgraded system, the up-to-date properties can be *verified at runtime* while executing the failing tests (Step 6) to discover the anomalous events that may explain the causes of *regression failures* (Step 7) [2, 11–14, 19]. In addition, the up-to-date properties can be *verified statically* to discover the *regression faults* that have not been revealed by any test case (Step 8) [15].

We applied this solution to multiple industrial systems, including a trajectory controller for robotic arms used in experimental nuclear reactors developed by VTT [17], a control system to detect spikes in large-scale power distribution networks developed by ABB [1], a real-time software framework for distributed control systems developed by ABB, and a controller for unmanned airborne vehicles developed by IAI [6]. Our experience with these systems revealed the presence of several important challenges that must be addressed to make runtime verification, and more specifically the analysis of regression problems, effective.

## 2   Challenges

This section presents the main challenges that we experienced when applying runtime verification to industrial software systems and describes the solutions that we experimented, as indicated in Table 1.

**Properties** We faced two main challenges related to properties: the lack of properties and the presence of inaccurate properties.

*Lack of Properties* The manual specification of the properties that can be verified at runtime is an *expensive* and *error-prone* activity. Engineers may deliberately put limited effort on the identification of the relevant program properties, producing very few properties, and thus drastically reducing the effectiveness of

Dynamic Analysis: Challenges and Possible Solutions      3

**Table 1.** Runtime Verification Challenges and Possible Solutions

| *Context* | *Challenge* | *Applied Solution* |
|---|---|---|
| Properties | Lack of properties | Derive properties automatically |
| | Inaccurate properties | Prune inaccurate properties |
| Monitoring | Impact of monitoring | Goal-driven monitoring |
| | Seamless Integration | Avoid code instrumentation |
| Applicability | Scalability of the analysis | Reduce the scope of the analysis |
| | Integration with the process | Exploit continuous integration |
| Effectiveness | Complex output | Organize the output hierarchically |
| | False positives | Priority-based filtering |

runtime verification. To overcome this issue, our idea is to automatically derive the program properties useful to detect regression problems by using *specification mining* techniques (Step 2 in Fig. 1).

So far we focused on the functional behaviour of the programs considering properties that capture the values of program variables (e.g., method pre- and post-conditions) [4] and the sequences of operations executed by a program (e.g., finite state machines) [9–11]. Results showed that mined properties could be a valid replacement of manually specified properties [11, 14, 15].

*Inaccurate Properties* Mined properties may overfit the data used for the mining, that is they capture properties that hold for the collected data but not for the program. Furthermore, in the presence of software changes, properties may become obsolete and be valid for the base version only. Overfitting and obsolete properties may generate false alarms, thus limiting the effectiveness of runtime verification and reducing the chance of its industrial adoption.

In order to be effective, our approach formally checks the correctness of mined properties (Step 3 in Fig. 1), and removes the properties that are intentionally invalidated by the changes (Step 5 in Fig. 1), thus *guaranteeing to work with up-to-date precise properties* when revealing failures and detecting faults [15].

**Monitoring** When monitoring industrial software systems, we faced two main challenges: keeping the impact of monitoring small and building a monitoring infrastructure easy to integrate with target systems.

*Impact of Monitoring* Monitoring slows down software systems, sometimes significantly altering the performance of an application and the outcome of the test cases. To address these issues we elaborated a *goal-driven monitoring* strategy that captures the strictly relevant events only.

Since it is intuitively true that incorrect changes are likely to affect the functionalities that directly depend on the changed code, we specifically restrict the monitoring to the neighbourhood of the changes [14]. At the same time we do not want to monitor the modified lines of code, because data collected from different software versions might be incomparable. Results show that the combination of these two criteria produces a focused monitoring strategy that may scale to industrial systems [14, 15].

*Seamless Monitoring Infrastructure* Monitoring code is often injected into applications using code transformation tools, which may require changes to the build process, such as changes to makefiles. These changes are hardly tolerated by developers, who are not happy to change program artefacts to enable specific analyses. We addressed this issue by only *using monitoring techniques that work with either the executables or the runtime environment.*

In particular, we implemented a monitoring tool that can generate GDB scripts [5] and PIN tools [7] with the capability to collect runtime data from the program functions that depend on the changes under analysis without requiring modifications to the build process. According to our experience, this approach is highly appreciated by software developers.

**Applicability** To make runtime verification easily applicable, it is important to define analysis strategies that both scale well with the complexity and size of industrial software systems and suitably integrate with tools for automation commonly available within professional organizations.

*Scalability of the Analysis* In presence of complex and large software systems, program analysis might not scale well. In our experience, this happened when using formal static analysis to prune the inaccurate properties mined from traces.

To deal with this issue, we had to find a *compromise between the soundness and the completeness* of the analysis. We thus decided to prune properties by running model checking on a *restricted scope* [15]. This might incidentally drop some true properties, but scales to large industrial systems. Findings compromises like this one is of crucial importance to address industrial systems.

*Integration with the Process* Industry people do not want to invest major effort on the configuration and the execution of tools. The adoption of novel analysis solutions, including runtime verification, might be facilitated if tools are implemented as components pluggable into *the environments commonly used in industry for automation.* In our experience, it has been a successful choice to develop our analysis technique as a Jenkins plugin [8] that can be simply installed in a continuous integration server.

**Effectiveness** To make runtime verification effective, it is important to produce outputs that can be easily inspected by the developers and to limit the number of false positives that can be generated.

*Complex Output* The explanation of a regression failure might depend on a number of mutually related events. In addition to identifying these events, it is important to present the information in a form that facilitates the inspection and the understanding of the failure.

We addressed this challenge by defining proper *views that show the cause-effect chains that relate anomalous events, and the hierarchical organization of the data* [2]. Compared to a plain list of apparently unrelated events, structured views are extremely easy to inspect.

*False positives* Techniques that generate *many false positives* are usually perceived as useless techniques by industry people. It is thus a priority of any analysis technique to limit the number of false positives that might be generated.

Since an analysis based on mined properties might produce false positives, we mitigated this issue by using *statically verified mined properties* [15] and by generating *structured views* that prioritize correct failure information to false positives, which usually are not inspected at all [2].

## 3 Conclusion

Our experience with industrial projects revealed the presence of several challenges that have to be faced to make runtime verification easily applicable to large and complex systems. In the context of evolving software, we identified strategies that can be used to face these challenges, specifically referring to functional faults. More work is needed to address a broader set of faults, (e.g. performance problems), and industrial systems, (e.g. low-power devices).

*Acknowledgments* This work has been partially supported by the H2020 Learn project, which has been funded under the ERC Consolidator Grant 2014 program (ERC Grant Agreement n. 646867).

## References

1. ABB. Power and automation company. http://www.abb.com/. Visited in 2016.
2. A. Babenko, L. Mariani, and F. Pastore. AVA: Automated interpretation of dynamically detected anomalies. In *ISSTA*. ACM, 2009.
3. H. Chockler, G. Denaro, M. Ling, G. Fedyukovich, A. E. J. Hyvrinen, L. Mariani, A. Muhammad, M. Oriol, A. Rajan, O. Sery, N. Sharygina, and M. Tautsching. Pincette - validating changes and upgrades in networked software. In *CSMR - EU Projects Track*. IEEE, 2013.
4. M. D. Ernst, J. Cockrell, W. G. Griswold, and D. Notkin. Dynamically discovering likely program invariants to support program evolution. *TSE*, 27(2):99–123, 2001.
5. FSF. Gdb debugger. http://sources.redhat.com/gdb/. Visited in 2016.
6. IAI. Israel aerospace industry. http://www.iai.co.il. Visited in 2016.
7. Intel. Pin - a dynamic binary instrumentation tool. https://software.intel.com/en-us/articles/pintool. Visited in 2016.
8. Jenkis. Continuous integration server. https://jenkins-ci.org/. Visited in 2016.
9. D. Lorenzoli, L. Mariani, and M. Pezzè. Automatic generation of software behavioral models. In *ICSE*. IEEE, 2008.
10. L. Mariani and F. Pastore. Automated identification of failure causes in system logs. In *ISSRE*. IEEE, 2008.
11. L. Mariani, F. Pastore, and M. Pezzè. Dynamic analysis for diagnosing integration faults. *IEEE TSE*, 37(4):486–508, 2011.
12. F. Pastore and L. Mariani. AVA: Supporting debugging with failure interpretations. In *ICST - Tool Demo Track*. IEEE, 2013.
13. F. Pastore, L. Mariani, and A. Goffi. Radar: a tool for debugging regression problems in C/C++ software. In *ICSE - Tool Demo Track*. IEEE, 2013.
14. F. Pastore., L. Mariani, A. Goffi, M. Oriol, and M. Wahler. Dynamic Analysis of Upgrades in C/C++ Software. In *ISSRE*. IEEE, 2012.
15. F. Pastore, L. Mariani, A. E. J. Hyvärinen, G. Fedyukovich, N. Sharygina, S. Sehestedt, and A. Muhammad. Verification-aided regression testing. In *ISSTA*. ACM, 2014.
16. G. Rothermel and M. J. Harrold. A safe, efficient regression test selection technique. *ACM TOSEM*, 6(2):173–210, 1997.
17. VTT. Research center. http://www.vtt.fi/. Visited in 2016.
18. K. Yu, M. Lin, J. Chen, and X. Zhang. Practical isolation of failure-inducing changes for debugging regression faults. In *ASE*. IEEE, 2012.
19. D. Zuddas, W. Jin, F. Pastore, L. Mariani, and A. Orso. Mimic: Locating and understanding bugs by analyzing mimicked executions. In *ASE*. ACM, 2014.